\documentclass[twocolumn,superscriptaddress,aps,prl]{revtex4-1}
\usepackage{amssymb,amsmath}
\usepackage{graphicx}
\usepackage[colorlinks,linkcolor=blue,citecolor=blue,urlcolor=blue]{hyperref}

\begin{document}

\title{Tuning a random field mechanism in a frustrated magnet}

\author{Shashikant Singh Kunwar}
\affiliation{Department of Physics, Indian Institute of Technology Madras, Chennai 600036, India.}

\author{Arnab Sen}
\affiliation{Department of Theoretical Physics, Indian Association for the Cultivation of Science, Jadavpur, Kolkata 700032, India.}

\author{Thomas Vojta}
\affiliation{Department of Physics, Missouri University of Science and Technology, Rolla, Missouri 65409, USA.}

\author{Rajesh Narayanan}
\affiliation{Department of Physics, Indian Institute of Technology Madras, Chennai 600036, India.}

\date{\today}

%\pacs{61.48.Gh,72.15.Rn, 73.22.Pr, 81.05.ue, 71.23.An, 71.23.-k, 71.55.Jv, 61.43.Hv, 64.60.al, 05.30.Rt}
%\keywords{ } %Use showkeys class option if keyword display desired

\begin{abstract}
We study the influence of spinless impurities on a frustrated magnet featuring a spin-density wave
(stripe) phase by means of Monte Carlo simulations.
We demonstrate that the interplay between the impurities and an order parameter that
breaks a real-space symmetry triggers the emergence of a random-field mechanism which destroys the
stripe-ordered phase. Importantly, the strength of the emerging random fields can be tuned by
the repulsion between the impurity atoms; they vanish for perfect anticorrelations between
neighboring impurities. This provides a novel way of controlling the phase diagram of a many-particle
system. In addition, we also investigate the effects of the impurities on the character of the
phase transitions between the stripe-ordered, ferromagnetic, and paramagnetic phases.
\end{abstract}

\maketitle

%%%%%%%%%%%%%%%%%%%%%%%%%%%%%%%%%%%%%%%%%%%%%%%%%%%%%%%%%%%%%%%%%%%%%%%%%%%%%%%%%%%%%%%%%%%%%%%%%%
\paragraph{Introduction:}
%%%%%%%%%%%%%%%%%%%%%%%%%%%%%%%%%%%%%%%%%%%%%%%%%%%%%%%%%%%%%%%%%%%%%%%%%%%%%%%%%%%%%%%%%%%%%%%%%%

Low-temperature phases of many-particle systems usually break one or several of the symmetries of
the interactions spontaneously. This is well described by the concept of order parameters (OPs),
quantities that vanish in the symmetric phase but are nonzero (and nonunique) in the symmetry-broken
phase (see, e.g., Ref.\ \cite{Goldenfeld}). A simple example of an OP is the total magnetization
which measures the degree to which the spin rotation symmetry is broken.
In recent years, lots of attention has been attracted by phases that spontaneously break real-space
symmetries in addition to spin, phase, or gauge symmetries, for example by rendering the $x$ and $y$
directions in a crystal inequivalent. Such phases include the charge-density wave or stripe phases in
cuprate superconductors, the Ising-nematic phases in the iron pnictides \cite{fradkin,fernandes,fkt15}, valence-bond-solids in quantum magnets~\cite{mlpm06, sandvik07, sen_sandvik10} and the crystalline phases of certain lattice-gas models of hard-core particles~\cite{rdd15}.

Realistic materials always contain some quenched disorder or randomness in the form of
vacancies, impurity atoms, random strains, and other types of imperfections. Consequently, the
question of how such randomness affects different broken symmetries and thus different OPs
is crucial for understanding the materials' behaviors (for recent reviews see, e.g., Refs.\ \cite{Vojta06,Vojta13}).

In this Letter, we focus on the impact of random disorder on a phase that breaks a real space symmetry.
To do so we turn our attention to a frustrated Ising model on a square lattice having ferromagnetic
nearest-neighbor interactions and antiferromagnetic next-nearest-neighbor interactions.
The disorder takes the form of spinless impurities or vacancies that dilute the magnetic lattice.
The resulting Hamiltonian reads
\begin{equation}
H = -J_{1}\sum_{\langle ij \rangle }\rho_{i}\rho_{j} S_{i}S_{j} - J_{2} \sum_{\langle \langle ij \rangle \rangle}\rho_{i}\rho_{j}S_{i}S_{j}
\label{eq:H}
\end{equation}
where the $S_{i} =\pm 1$ are classical Ising variables, while $J_1>0$ and $J_2<0$ are the nearest-neighbor and
next-nearest-neighbor interactions, respectively. The $\rho_i$ are quenched random variables that take
the values 0 (vacancy) with probability $p$ and $1$ (site occupied by spin) with probability $1-p$.
We consider both uncorrelated randomness for which the $\rho_i$ are statistically independent
and anticorrelated randomness for which repulsion between the impurities suppresses the simultaneous
occupation of two nearest-neighbor sites by impurities.

In the absence of vacancies ($p=0$), the phase diagram and the phase transitions of this system are
well-understood (see, e.g., Refs.\ \cite{arnab1,arnab2,kalz,kalzprb2012} and references therein). At high temperatures,
it features a conventional paramagnetic phase. Upon lowering the temperature, two distinct symmetry-broken
phases appear. For $g=|\rm{J_2}|/\rm{J_1} < 1/2$, the system enters a ferromagnetic (FM) low-temperature phase
that breaks the $Z_2$ Ising symmetry but none of the real-space symmetries. For $g>1/2$, in contrast, the low-temperature
phase displays a stripe-like spin order that breaks not only the Ising symmetry but also the $Z_4$
rotation symmetry of the square lattice. The Hamiltonian (\ref{eq:H}) is thus particulary
well suited for our study as it allows us to contrast an OP that does not break any real-space
symmetries with one that does.

\begin{figure}
\includegraphics[scale=0.45,angle=0]{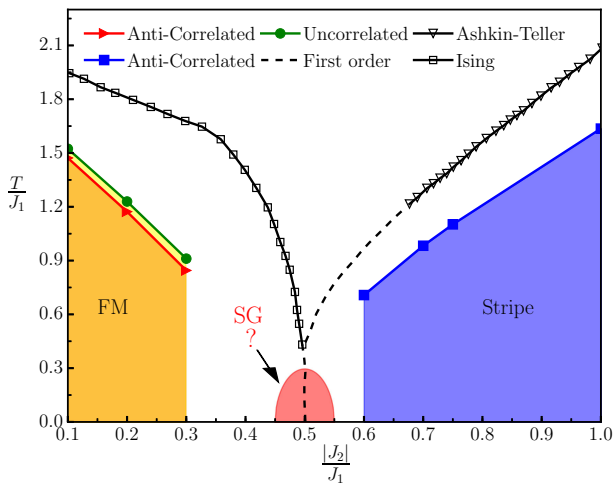}
\caption{Phase diagram of $\rm{J_1}$-$\rm{J_2}$ Hamiltonian (\ref{eq:H}) for both uncorrelated and anti-correlated site dilution at an impurity concentration of $p=1/8$ compared to the phase diagram of the undiluted system (open symbols) \cite{kalz,kalzprb2012}. For  uncorrelated impurities, the emergent random field mechanism destroys the stripe-ordered phase.
In contrast, this phase survives the introduction of anti-correlated disorder. }
\label{fig:fig1}
\end{figure}

To analyze how the site dilution influences the frustrated Ising model (\ref{eq:H}), we perform extensive Monte Carlo simulations.
We also determine the exact ground states of small plaquettes to understand the disorder effects microscopically.
Our results are illustrated by the phase diagram shown in Fig.\ \ref{fig:fig1} and
can be summarized as follows.
The ferromagnetic low-temperature phase survives moderate dilution with both uncorrelated and anticorrelated impurities, 
but its Curie temperature $T_c$ is suppressed. In contrast, the stripe-ordered low-temperature phase is completely absent for 
uncorrelated impurities. This is caused by an effective random field for the stripe order that emerges due to the interplay 
of the impurities and the broken real-space symmetry. This emergent random field destroys the stripe order via domain 
formation \cite{imryma,fernandez}. Importantly, the strength of the random fields can be controlled by the repulsion between 
the impurities; it completely vanishes if the repulsion prohibits the simultaneous occupation of nearest-neighbor sites by impurities. 
In this case of perfect local anticorrelations between the impurities, the stripe-ordered low-temperature phase survives, albeit with a depressed critical temperature $T_c$ compared to the
undiluted system. This tunable random-field mechanism is the main result of this Letter. In addition, we demonstrate that the first-order phase transitions of the undiluted system are rounded by the disorder, in line with the Aizenmann-Wehr theorem \cite{aizenmann, berker}.
In the rest of this Letter, we discuss our simulations, explain the tunable random-field mechanism, and put our results
into a broader perspective.

%%%%%%%%%%%%%%%%%%%%%%%%%%%%%%%%%%%%%%%%%%%%%%%%%%%%%%%%%%%%%%%%%%%%%%%%%%%%%%%%%%%%%%%%%%%%%%%%%%
\paragraph{Monte Carlo simulations:}
%%%%%%%%%%%%%%%%%%%%%%%%%%%%%%%%%%%%%%%%%%%%%%%%%%%%%%%%%%%%%%%%%%%%%%%%%%%%%%%%%%%%%%%%%%%%%%%%%%

We employ standard single-spin flip Metropolis \cite{MRRT53}  simulations of the Hamiltonian
(\ref{eq:H}). We study square lattices of linear sizes between $L=8$ to 80, averaging the results over
500 to 1000 disorder configurations. Details of the simulation algorithm and parameter values can be found in the
Supplemental Material \cite{Supplemental}.
The primary observables are the OPs for the ferromagnetic and stripe phases.
The two-component stripe OP $\psi \equiv (\psi_x,\psi_y)$ is defined as \cite{arnab1,arnab2}
\begin{equation}
\psi_x = \frac{1}{L^{2}}\sum_{i}\rho_{i}S_i(-1)^{x_i}, \quad \psi_y = \frac{1}{L^2}\sum_{i=1}\rho_{i}S_i(-1)^{y_i},
\label{eq:stripe_OP}
\end{equation}
where $(x_i,y_i)$ are the coordinates of site $i$ whereas the ferromagnetic OP, i.e., the magnetization,
reads
\begin{equation}
m = \frac{1}{L^{2}}\sum_{i}\rho_{i}S_i~.
\label{eq:FM_OP}
\end{equation}
We also analyze the corresponding susceptibilities
$\chi_S = L^{2}\left[\langle \psi^2\rangle - \langle |\psi |\rangle^2\right]/T$ and
$\chi_F = L^{2}\left[\langle m^2\rangle - \langle |m |\rangle^2\right]/T$ as well as
the Binder cumulants
\begin{equation}
U_{\rm S} = 2 \left( 1-\frac{1}{2}\frac{\left[\langle \psi^4 \rangle\right]}{{\left[\langle \psi^2 \rangle\right]}^2}\right), ~
U_{\rm F} = \frac 3 2 \left( 1-\frac{1}{3}\frac{\left[\langle m^4 \rangle\right]}{{\left[\langle m^2 \rangle\right]}^2}\right)~.
\label{eq:Binder}
\end{equation}
Here, $\left[\cdots\right]$ denotes the average over disorder realizations whereas $\langle\cdots\rangle$ indicates the usual
thermodynamic (Monte Carlo) average. The Binder cumulants are normalized such that they take the limiting values
$U_{\rm F,S} \rightarrow 1$ deep in the corresponding ordered phases and  $U_{\rm F,S} \rightarrow 0$ deep in the disordered phase.
The crossing of the Binder cumulant curves for different system sizes yields the location of the phase transition.
The Binder cumulant also allows us to determine the order of the transition: For a continuous transition, it is a monotonic function of temperature \cite{binder}. At a first-order transition, in contrast, the Binder cumulant shows a minimum that becomes more pronounced with increasing system size \cite{vollmayr} and is caused by the existence of multiple peaks in the OP distribution. This non-monotonic
 temperature dependence can serve as an indicator of a first-order transition.

%%%%%%%%%%%%%%%%%%%%%%%%%%%%%%%%%%%%%%%%%%%%%%%%%%%%%%%%%%%%%%%%%%%%%%%%%%%%%%%%%%%%%%%%%%%%%%%%%%
\paragraph{Stripe Phase:}
%%%%%%%%%%%%%%%%%%%%%%%%%%%%%%%%%%%%%%%%%%%%%%%%%%%%%%%%%%%%%%%%%%%%%%%%%%%%%%%%%%%%%%%%%%%%%%%%%%

\begin{figure}
\includegraphics[scale=0.42]{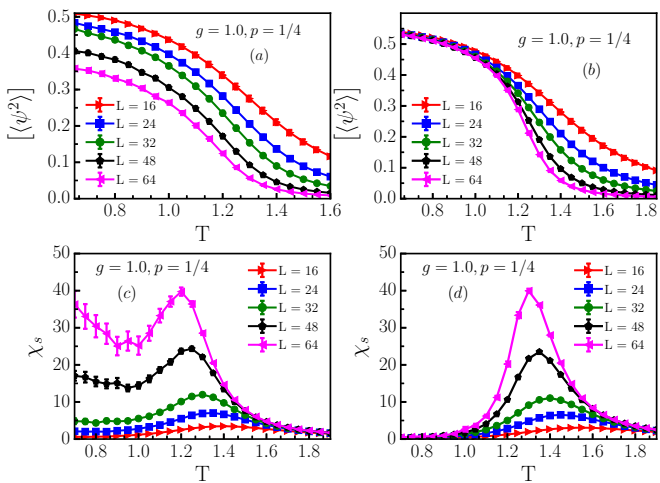}
\caption{Stripe order-parameter $\psi$ and stripe susceptibility $\chi_S$  as functions of temperature $T$ for frustration parameter
$g=1$, dilution $p=1/4$ and several system sizes. Data for uncorrelated vacancies are shown in panels (a) and (c) whereas
panels (b) and (d) show results for anti-correlated vacancies.}
\label{fig:stripeop}
\end{figure}

We now turn to the central question of this Letter, the fate of the stripe phase upon introducing
spinless impurities. Figure \ref{fig:stripeop} depicts the stripe OP and the associated
susceptibility for dilution $p=1/4$, contrasting the cases of uncorrelated impurities and perfectly
anticorrelated impurities (where the simultaneous occupation of nearest-neighbor sites by impurities is forbidden).
The frustration parameter is $g=|J_2|/J_1=1$ for which the undiluted system
features a stripe-ordered low-temperature phase.
Figure \ref{fig:stripeop}(a) shows that the stripe order-parameter at low temperatures decreases with increasing system size
for the case of uncorrelated impurities. In this case, the stripe susceptibility shown in Fig.~\ref{fig:stripeop}(c)
develops a pronounced secondary peak at low temperatures.  As suggested in Ref.~\cite{fernandez}, these observations
indicate the absence of long-range stripe order in the thermodynamic limit. In contrast, in the case of anticorrelated disorder, the stripe order-parameter saturates at a size-independent nonzero value at low temperatures, as shown in Fig.~\ref{fig:stripeop}(b).
The corresponding stripe susceptibility, shown in Fig.~\ref{fig:stripeop}(d), displays the conventional behavior associated with
a continuous phase transition. These observations suggest that the stripe order survives in the case of anticorrelated impurities.

\begin{figure}
\includegraphics[scale=0.42,angle=-0]{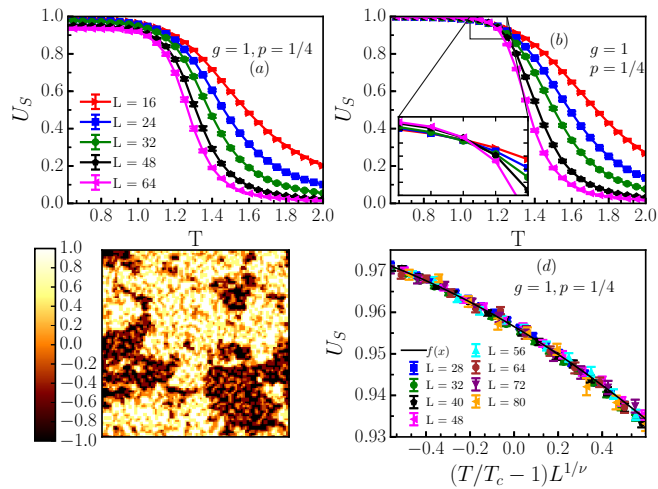}
\caption{Stripe Binder cumulant $U_S$ vs.\ temperature $T$ for frustration parameter
$g=1$, dilution $p=1/4$ and several system sizes. Panel (a) shows data for uncorrelated impurities
whereas results for anticorrelated impurities are presented in panel (b). Panel (c): Local nematic
OP $\eta_i$ for a single system of 100x100 sites, uncorrelated impurities with
dilution $p=1/4$, $T=0.55$, and $g=1$.
Panel (d) shows the scaling collapse (with $\bar{\chi}^2 =0.97$)  of the stripe Binder cumulant for
anticorrelated impurities and $g=1$, $p=1/4$.}
\label{fig:US}
\end{figure}

To provide further evidence, we compare the behavior of the stripe Binder cumulants $U_{\rm S}(T)$ for
uncorrelated and anticorrelated impurities. Figure \ref{fig:US} depicts the Binder cumulants for the same parameters used above,
viz., $p=1/4$ and $g=1$.
Focussing on Fig.~\ref{fig:US}(a), we see that for uncorrelated impurities, the Binder cumulant vs.\ temperature curves
for different system sizes do not cross. With increasing size, the Binder cumulant shifts to smaller and smaller values, i.e., towards the
disordered phase, confirming the absence of long-range stripe order for the case of uncorrelated
impurities. The fate of the stripe phase can be further illustrated via the nematic OP $\eta= \psi_x^2 -\psi_y^2$
which measures the local preference for vertical vs. horizontal stripes. The color plot in Fig.~\ref{fig:US}(c) shows
the \emph{local} nematic OP for each $2\times 2$ plaquette, clearly demonstrating
competing domains of horizontal and vertical stripes \cite{Supplemental}.

In contrast, for the case of anticorrelated impurities, the stripe Binder cumulants for different system sizes do cross
as evidenced in Fig.~\ref{fig:US}(b). This indicates the existence of a phase transitions and thus the survival of the stripe-ordered
low-temperature phase. Estimates of the transition temperature $T_c$ and the correlation length exponent $\nu$ can be obtained
from finite-size scaling \cite{Barber_review83,Cardy_book88} (for details, see Supplemental material \cite{Supplemental}).
Figure \ref{fig:US}(d) shows the scaling collapse of the Binder cumulant
in terms of the scaled variable $(T-T_c)L^{1/\nu}$, with $T_c=1.1729(5)$ and $\nu= 1.26(3)$.
The data collapse is very good; the underlying least-square
fit has a reduced $\bar{\chi}^2 =0.97$ \cite{note1}. 
 Because our systems are only moderately large,
the value of $\nu$ should be understood as an effective exponent rather than the true asymptotic exponent.

%%%%%%%%%%%%%%%%%%%%%%%%%%%%%%%%%%%%%%%%%%%%%%%%%%%%%%%%%%%%%%%%%%%%%%%%%%%%%%%%%%%%%%%%%%%%%%%%%%
\paragraph{Random fields from spinless impurities:}
%%%%%%%%%%%%%%%%%%%%%%%%%%%%%%%%%%%%%%%%%%%%%%%%%%%%%%%%%%%%%%%%%%%%%%%%%%%%%%%%%%%%%%%%%%%%%%%%%%

\begin{figure}
\includegraphics[scale=0.66,angle=0]{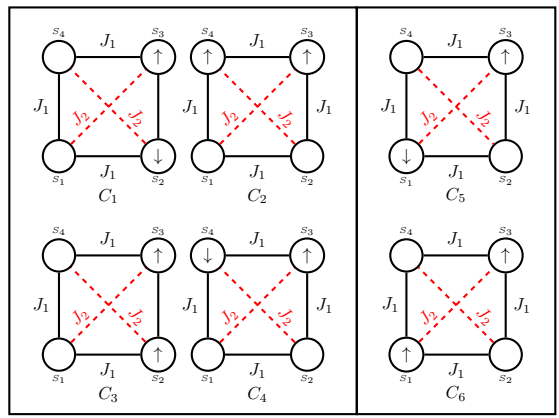}
\caption{Impurity configurations on $2\times 2$ plaquettes illustrating the emergence
of random-field disorder for the stripe OP (see text for further details).}
\label{fig:plaquettes}
\end{figure}

To explain the absence of the stripe phase for uncorrelated impurities, we now demonstrate that
the impurities induce effective random fields for the nematic OP $\eta= \psi_x^2 -\psi_y^2$.
We focus on the ground state energies of small
plaquettes of $2 \times 2 $ sites as seen in Fig.\ \ref{fig:plaquettes}.
If impurities simultaneously occupy two vertical nearest-neighbor sites (configurations $C_1$ and $C_3$ in Fig.\ \ref{fig:plaquettes}),
vertical stripes (configuration $C_3$) are favored over horizontal stripes (configuration $C_1$)
as their ground state energy on the plaquette is lower by $-2J_1$. Analogously, if impurities
occupy two horizontal nearest-neighbor sites (configurations $C_2$ and $C_4$), horizontal stripes ($C_2$) are favored
over vertical stripes ($C_4$). In contrast, configurations with either a single impurity
or two impurities across the diagonal of a plaquette ($C_5$ and $C_6$) do not prefer one stripe orientation
over the other.

This means that impurity configurations in which two impurities occupy nearest neighbor sites
\emph{locally} break the $Z_4$ lattice rotation symmetry. They thus act as random fields
for the nematic OP $\eta$ by locally preferring either the $\psi_x$ or the $\psi_y$ component of the
stripe OP (\ref{eq:stripe_OP}). As was argued by Imry and Ma \cite{imryma} in the
context of the random-field Ising model \cite{nattermann1} and later proven
rigorously \cite{aizenmann}, random fields destroy the long-range ordered phase via domain formation.
Monte Carlo evidence for domains was presented in Fig.\ \ref{fig:US}(c).

The typical size $L_D$ of these domains depends on the strength of the random fields and thus on the
dilution $p$. In two dimensions, the dependence is expected to be exponential,
$L_D \sim \exp \left({\rm const}/p^4\right)$, for small $p$ \cite{nattermann1}.
This implies that the domain size will exceed the system size for sufficiently small $p$,
making the destruction of the long-range order unobservable \cite{Supplemental}.

The fact that a local preference for vertical or horizontal stripes only appears if two impurities
occupy two nearest-neighbor sites can be used to tune the strength of the emerging random field mechanism.
If the probability for nearest-neighbor pairs of impurities is reduced, for example because
of a repulsive interaction between the impurities, fewer random fields appear in the system.
In the limit of perfectly anticorrelated impurities where such pairs are completely forbidden,
the random-field mechanism is switched off \cite{note2}.
This explains why our simulations showed that the stripe-ordered phase survives for anticorrelated
impurities.

%%%%%%%%%%%%%%%%%%%%%%%%%%%%%%%%%%%%%%%%%%%%%%%%%%%%%%%%%%%%%%%%%%%%%%%%%%%%%%%%%%%%%%%%%%%%%%%%%%
\paragraph{Ferromagnetic phase:}
%%%%%%%%%%%%%%%%%%%%%%%%%%%%%%%%%%%%%%%%%%%%%%%%%%%%%%%%%%%%%%%%%%%%%%%%%%%%%%%%%%%%%%%%%%%%%%%%%%

In contrast to the stripe OP, the total magnetization
does not break a real-space symmetry. Therefore, spinless impurities do not create random fields
coupling to the ferromagnetic order. Instead, they act as much more benign random-mass or random-$T_c$
disorder. Consequently, the ferromagnetic phase survives in the presence of impurities, be they
uncorrelated or perfectly anticorrelated. However, the Curie temperature $T_c$  is reduced compared
to the undiluted system, as is shown in the phase diagram in Fig.\ \ref{fig:fig1}.

%%%%%%%%%%%%%%%%%%%%%%%%%%%%%%%%%%%%%%%%%%%%%%%%%%%%%%%%%%%%%%%%%%%%%%%%%%%%%%%%%%%%%%%%%%%%%%%%%%
\paragraph{Phase transitions:}
%%%%%%%%%%%%%%%%%%%%%%%%%%%%%%%%%%%%%%%%%%%%%%%%%%%%%%%%%%%%%%%%%%%%%%%%%%%%%%%%%%%%%%%%%%%%%%%%%%

We now turn to the phase transitions between the paramagnetic, ferromagnetic, and stripe phases.
The transitions of the undiluted system are well understood \cite{arnab1,arnab2,kalz,kalzprb2012}.
As illustrated in Fig.\ \ref{fig:fig1},  there is a direct first-order
phase transition between the ferromagnetic and stripe phases at low temperatures. The transition between
the ferromagnetic and paramagnetic phases is continuous and
belongs to the 2D Ising universality class. Extensive numerical simulations
have also established that the transition from the stripe phase to
the paramagnetic phase is of first order for  $g<g^* \approx 0.67$.
The line of first-order transition terminates at $g^*$ and gives rise to
critical behavior that belongs to the Ashkin-Teller universality class
\cite{AToriginal}.

In the presence of anticorrelated disorder, the ferromagnetic and stripe phases both survive.
According to Landau \cite{Landau}, phase transitions between two ordered phases that break different
symmetries must be of first order. However, the Aizenman-Wehr theorem \cite{aizenmann} forbids
first-order transitions in two-dimensional disordered systems. This implies that the ferromagnetic
and stripe phases must be separated by an intermediate phase. This could simply be the
paramagnetic phase extending all the way to zero temperature, or there could be a spin glass (SG) phase 
at low temperatures and $g$ close to 0.5. Unequivocally resolving the phases in this parameter region
is beyond the scope of this Letter.

\begin{figure}
\includegraphics[scale=0.43,angle=-0]{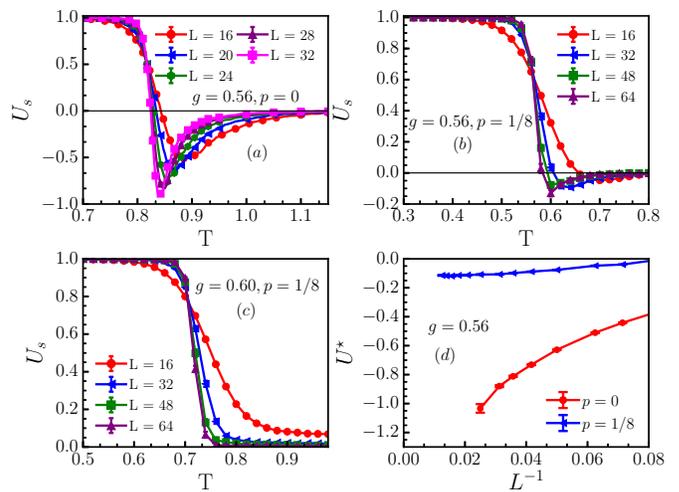}
\caption{Stripe Binder cumulant $U_S$ vs.\ temperature for different system sizes.
(a) undiluted system, $p=0$, $g=0.56$; (b) anticorrelated impurities, $p=1/8$, $g=0.56$;
(c) anticorrelated impurities, $p=1/8$, $g=0.60$; (d) minimum value $U^*$ as a function
of inverse system size.}
\label{fig:fig6}
\end{figure}

The stripe to paramagnetic transition of the undiluted system is of first-order for
$0.5 < g < g^*\approx 0.67$. To determine the character of this transition in the presence
of anticorrelated impurities, we analyze the stripe Binder cumulant $U_S$ in Fig.\ \ref{fig:fig6}.
In the undiluted system depicted in Fig.\ \ref{fig:fig6}(a), $U_S$ shows a
pronounced minimum close to the transition which gets deeper with system size (see also
Fig.\ \ref{fig:fig6}(d)). This clearly indicates a first-order transition. In contrast, in
the diluted system with $p=1/8$ and $g=0.6$ shown in Fig.\ \ref{fig:fig6}(c),
$U_S$ does not feature any minima,
demonstrating that the first-order transition is rounded to a continuous one, in agreement with the
Aizenman-Wehr theorem \cite{aizenmann}. For the diluted system at $g=0.56$, the Binder cumulant
shows weak minima, but they do not deepen with system size. This can be attributed to the fact that
the clean first-order transition is stronger at smaller $g$. The disorder-induced rounding will
therefore occur at a larger length scale beyond the moderate sizes used in our simulations.
This is compatible with the size dependence shown in Fig.\ \ref{fig:fig6}(d).

The ferromagnetic to paramagnetic transition survives for both uncorrelated and anticorrelated
impurities. The critical behavior across all phase transition lines in the diluted case is compatible
with the two-dimensional Ising universality class with logarithmic corrections, as is
discussed in the Supplemental Material \cite{Supplemental}.

%%%%%%%%%%%%%%%%%%%%%%%%%%%%%%%%%%%%%%%%%%%%%%%%%%%%%%%%%%%%%%%%%%%%%%%%%%%%%%%%%%%%%%%%%%%%%%%%%%
\paragraph{Conclusions:}
%%%%%%%%%%%%%%%%%%%%%%%%%%%%%%%%%%%%%%%%%%%%%%%%%%%%%%%%%%%%%%%%%%%%%%%%%%%%%%%%%%%%%%%%%%%%%%%%%%

In summary, we have studied the effects of spinless impurities on the phases of a frustrated Ising magnet.
As the impurities do not break the Ising symmetry of the ferromagnetic OP, they act as
rather benign random-mass disorder in the ferromagnetic phase. Consequently, this phase survives in the
presence of the impurities, albeit with reduced Curie temperature. In contrast, the impurities can
locally break the symmetry between horizontal and vertical stripes and thus create effective random
fields for the nematic OP. These emerging random fields destroy the stripe phase via
domain formation.

The microscopic understanding of the random fields has allowed us to identify a way to
tune their strength. The random fields are suppressed with increasing repulsion
between the impurities and completely vanish if nearest-neighbor pairs of impurities are forbidden.
Therefore, the stripe phase survives for such perfectly anticorrelated impurities.
This mechanism offers a novel way of controlling the phase diagram of a many-particle system.
Note that the protection of the stripe phase by local (anti-)correlations between the impurities is similar to the protection
of a clean quantum critical point by local disorder correlations discussed in Ref.\
\cite{joselaflorencieetal}.

Finally, we comment on the possibility of a nematic phase in the Hamiltonian (\ref{eq:H}). In principle,
the paramagnetic to stripe phase transition could split into two separate transitions:
The $Z_4$ lattice symmetry is broken first, leading to nematic order,
while the Ising spin symmetry is broken at a lower temperature. Nematic order has indeed
been observed in a $J_1$-$J_2$ model in an external field \cite{alejandra}. However, our
simulations have not provided any indications of a nematic phase in our problem.

%%%%%%%%%%%%%%%%%%%%%%%%%%%%%%%%%%%%%%%%%%%%%%%%%%%%%%%%%%%%%%%%%%%%%%%%%%%%%%%%%%%%%%%%%%%%%%%%%%
\paragraph{Acknowledgements:}
%%%%%%%%%%%%%%%%%%%%%%%%%%%%%%%%%%%%%%%%%%%%%%%%%%%%%%%%%%%%%%%%%%%%%%%%%%%%%%%%%%%%%%%%%%%%%%%%%%

A.S. is partly supported through the Partner Group program between the Indian Association for the Cultivation of Science (IACS), Kolkata and the Max Planck Institute for the Physics of Complex Systems (MPIPKS), Dresden. T.V. is supported in part by the NSF under
Grants No. PHY-1125915 and No. DMR- 1506152.
The numerical data were generated at the PG Senapathy computing facility at Indian Institute of Technology, Madras, the computing facility at the Department of Theoretical Physics, IACS and the computing facility at MPIPKS.

%%%%%%%%%%%%%%%%%%%%%%%%%%%%%%%%%%%%%%%%%%%%%%%%%%%%%%%%%%%%%%%%%%%%

\setcounter{equation}{0}
\setcounter{figure}{0}
\setcounter{table}{0}
\setcounter{page}{1}
\makeatletter
\renewcommand{\theequation}{S\arabic{equation}}
\renewcommand{\thefigure}{S\arabic{figure}}
\renewcommand{\bibnumfmt}[1]{[S#1]}
\renewcommand{\citenumfont}[1]{S#1}
\newpage
%%%%%%%5
%SUPINFO
%%%%%%%
\makeatother
\clearpage 
\onecolumngrid
%%%%%%%%%%%%%%%%%%%%%%%%%%%%%%%%%%%%%%%%%%%%%%%%%%%%%%%%%%%%%%%%%%%%%%%%%%%%%%%%%%%%%%%%%%%%
{\LARGE\bf \noindent 
Supplemental material for:\\[0.5ex] Tuning a random field mechanism in a frustrated magnet
}
\bigskip

\noindent 
Shashikant Singh Kunwar$^1$, Arnab Sen$^2$, Thomas Vojta$^3$, and Rajesh Narayanan$^1$
\bigskip

{\small \noindent
$^1$ Department of Physics, Indian Institute of Technology Madras, Chennai 600036, India.\\
$^2$ Department of Theoretical Physics, Indian Association for the Cultivation of Science, Jadavpur, Kolkata 700032, India.\\
$^3$ Department of Physics, Missouri University of Science and Technology, Rolla, Missouri 65409, USA.
}
%
%%%%%%%%%%%%%%%%%%%%%%%%%%%%%%%%%%%%%%%%%%%%%%%%%%%%%%%%%%%%%%%%%%%%%%%%%%%%%%%%%%%%%%%%%%%%%%%%%%%%%%
\section*{S1. Details of Monte-Carlo procedure}
%%%%%%%%%%%%%%%%%%%%%%%%%%%%%%%%%%%%%%%%%%%%%%%%%%%%%%%%%%%%%%%%%%%%%%%%%%%%%%%%%%%%%%%%%%%%%%%%%%%%%%
We employ the classical single-spin-flip Metropolis algorithm \cite{S_MRRT53} to perform our simulations
because cluster-flip methods such as the Swendsen-Wang \cite{S_SwendsenWang87} and Wolff \cite{S_Wolff89}
algorithms do not improve the performance in the presence of frustrated interactions. We study square
lattices of linear size $L=8$ to 80. Each Monte Carlo simulation consists of an equilibration period
of $10^6$ Monte Carlo sweeps (a sweep corresponds to one attempted spin flip per lattice site), 
followed by a measurement period of another $10^6$ sweeps, with measurements taken after each sweep.
To improve the equilibration performance, we adopt a cooling procedure. We start the simulations
at high temperatures and lower the temperature in small steps, using the final state of the higher 
temperature simulation as the initial condition for the next lower temperature.
 
We investigate frustration parameters $g=|J_2|/J_1$ between 0.1 and 1.0. To study the influence of disorder, 
a total number of $N_{imp} = p L^2$ spinless impurity sites are introduced
into the lattice. These impurities are either completely uncorrelated or they are perfectly anticorrelated such 
that the simultaneous occupation of nearest-neighbor sites by impurities is forbidden. We simulate dilutions
of $p=1/8$ and 1/4. We expect, however, that the qualitative results hold for all values of $p$ that are
sufficiently small such that lattice percolation effects do not play a role. All observables are
averaged over $1000$ impurity configurations for the smaller system sizes, $L=8$ to $32$, and over 500
impurity configurations for the larger sizes.

%%%%%%%%%%%%%%%%%%%%%%%%%%%%%%%%%%%%%%%%%%%%%%%%%%%%%%%%%%%%%%%%%%%%%%%%%%%%%%%%%%%%%%%%%%%%%%%%%%%%%%
\section*{S2. Finite-size scaling analysis}
%%%%%%%%%%%%%%%%%%%%%%%%%%%%%%%%%%%%%%%%%%%%%%%%%%%%%%%%%%%%%%%%%%%%%%%%%%%%%%%%%%%%%%%%%%%%%%%%%%%%%%

In this section we describe  the methodology adopted to extract the critical temperature $T_c$ and the 
critical exponents from the Monte Carlo data of the site-diluted ${J_1}$-${J_2}$ model.
The analysis is based on finite-size scaling \cite{S_Barber_review83,S_Cardy_book88} of the stripe and 
ferromagnetic Binder cumulants $U_S$ and 
$U_F$ as well as the corresponding susceptibilities $\chi_S$ and $\chi_F$.

\subsection*{S2.1 Ferromagnetic transition}

\begin{figure}
\centerline{\includegraphics[width=0.75\textwidth]{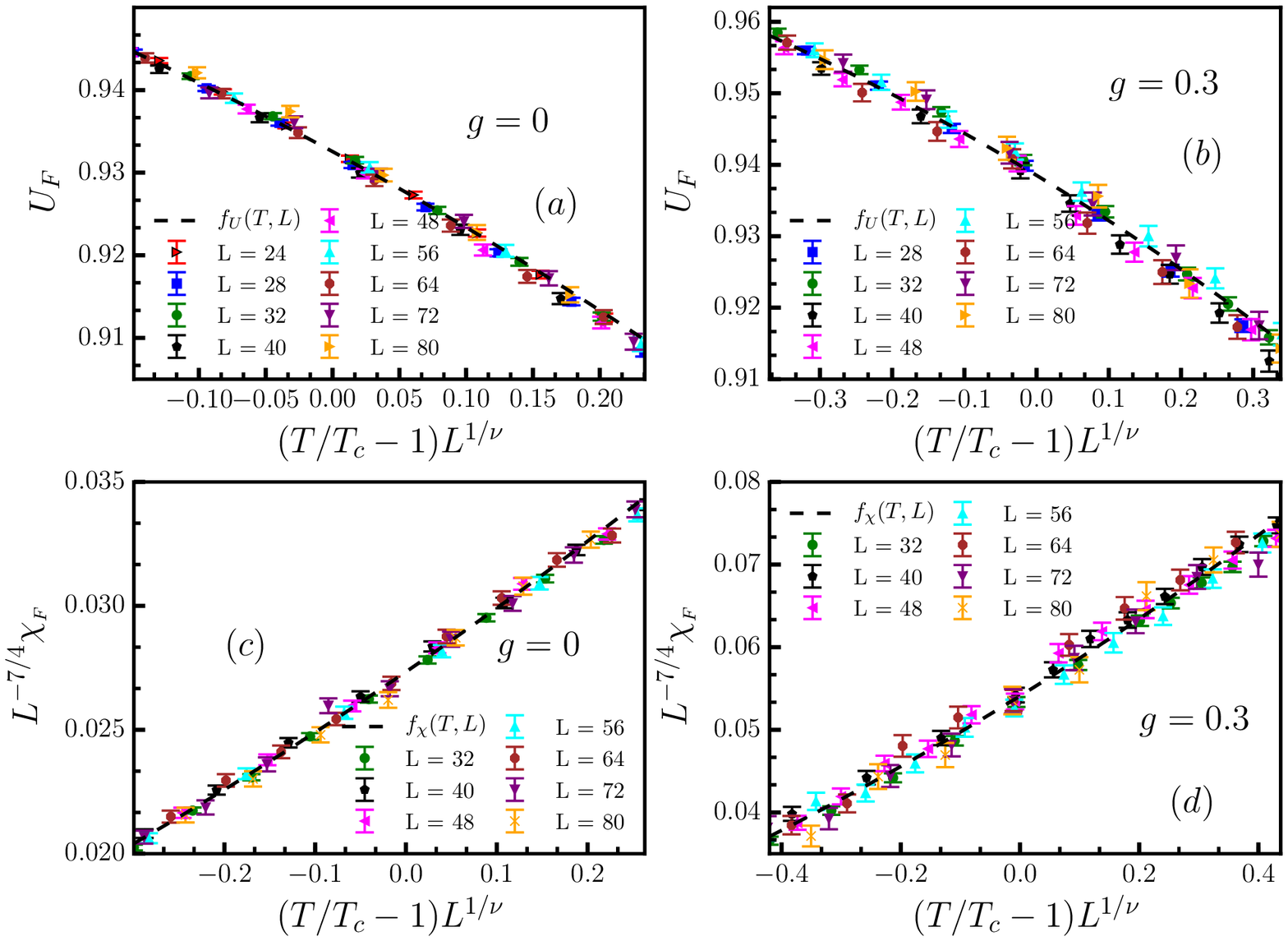}}
\caption{Scaling collapse of the ferromagnetic binder cumulant $U_{F}$ [panels (a) and (b)]
and the scaled ferromagnetic susceptibility $\chi_{F}L^{-7/4}$ [panels (c) and (d)] for
uncorrelated impurities at $p=1/8$ and frustration parameters $g=0$ and 0.3.}
\label{scalferrornd}
\end{figure}

We start by analyzing the ferromagnetic Binder cumulant $U_{\rm F}$ defined as
\begin{equation}
U_{\rm F} = \frac 3 2 \left( 1-\frac{1}{3}\frac{\left[\langle m^4 \rangle\right]}{{\left[\langle m^2 \rangle\right]}^2}\right)~.
\label{eq:Binder_F}
\end{equation}
According to finite-size scaling, the Binder cumulant values for different system sizes $L$ and temperatures $T$
should collapse onto a single master curve when plotted as a function of the scaling variable $x=(T-T_c) L^{1/\nu}$
where $\nu$ is the correlation length critical exponent.
Moreover, as the Binder cumulant is a dimensionless quantity, its value right at $T_c$ should be size-independent,
implying a Taylor expansion
\begin{equation}
\label{eq:binfit}
U_{\rm F,S}(T,L) = f(x) = a_{0} + a_{1} x + a_{2} x^2 + \ldots.
\end{equation}
sufficiently close to the critical point.
Figures \ref{scalferrornd}(a) and (b) show examples of such scaling plots for
uncorrelated impurities at concentration $p=1/8$ and frustration parameters $g=0$ and 0.3, respectively.
The values of $T_c$ and $\nu$ are extracted from fits of the $U_F$ data to the expansion (\ref{eq:binfit})
truncated after the quadratic term.
The quality of the fit can be estimated from the reduced sum of squared errors (per degree of freedom) $ \bar{\chi}^{2}$ defined as
\begin{equation}\label{chi2def}
\bar{\chi}^{2}  = \frac{1}{N-M} \sum_{i = 1}^{N}\frac{[U_{F,i} - f(x_{i})]^{2}}{\sigma_{i}^{2}}~.
\end{equation}
Here, $N$ is the number of data points, $M$ is the number of fit-parameters, and $\sigma_{i}^2$ is the (Monte Carlo)
variance of the value $U_{F,i}$. The fits are considered of good quality when $\bar{\chi}^{2} \lessapprox 1$.
Results of this analysis for both uncorrelated and anticorrelated impurities and several values
of the frustration parameter $g$ are presented in Table \ref{finaltable1}.
\begin{table}
\renewcommand*{\arraystretch}{1.2} 
\begin{center}
\begin{tabular*}{13cm}{c |@{\extracolsep{\fill}} ccc|ccc}
\hline\hline
  ~  &\multicolumn{3}{c}{anticorrelated} & \multicolumn{3}{c}{uncorrelated} \\
$g$ &$T_{c}$ &$\nu$ & $\bar{\chi}^{2}$  & $T_{c}$ & $\nu$ & $\bar{\chi}^{2}$  \\
\hline
$0$ &$1.7574(1)$&$1.16(1)$& $1.19$  &$1.8036(1)$ &$1.12(1)$& $1.36$ \\
$0.1$&$1.4724(1)$&$1.11(1)$& $0.71$ &$1.5234(1)$ &$1.13(2)$ & $1.37$  \\
$0.2$&$1.1728(1)$&$1.14(3)$& $1.01$  &$1.2294(1)$ &$1.17(2)$ & $1.10$ \\
$0.3$ &$0.8450(2)$&$1.14(4)$& $0.82$ &$0.9108(2)$ &$1.15(4)$ & $1.45$  \\
\hline\hline
\end{tabular*}
\caption{Critical temperatures $T_c$, effective correlation length exponents $\nu$, and reduced error sums $\bar\chi^2$ 
obtained from the scaling analysis of the ferromagnetic Binder cumulant $U_F$.  
Results are shown for various values of the frustration parameter $g$ and dilution $p=1/8$
for both uncorrelated and anticorrelated impurities. The numbers in parentheses give the
error of the last digit.}
\label{finaltable1}
\end{center}
\end{table}

How do our results for the correlation length exponent $\nu$ compare to theoretical predictions?
The ferromagnetic-to-paramagnetic transition in the clean, undiluted system belongs to the two-dimensional
Ising universality class. Its correlation length exponent takes the value $\nu_{cl}=1$ implying that 
random-mass disorder is exactly marginal according to the Harris criterion $d\nu >2$ \cite{S_harris}. 
The fate of the phase transition in the two-dimensional disordered Ising model has been controversially discussed
in the literature (see, e.g., Ref.\ \cite{S_qiong} and references therein). 
Recent numerical results \cite{S_qiong} demonstrate, however, that the critical behavior of the disordered
Ising model is controlled
by the \emph{clean} two-dimensional Ising critical point but with universal logarithmic corrections
as predicted by perturbative renormalization group calculations.
Our system sizes are too small to reliably extract logarithmic corrections. The $\nu$ values in
Table \ref{finaltable1} must therefore be considered effective rather than asymptotic exponent
values. They are comparable to effective $\nu$ values found in the above-mentioned high-precision
study of the disordered Ising model. We thus conclude that our results
are consistent with the critical behavior of the ferromagnetic transition belonging to the disordered 
Ising universality class.

Further evidence is provided by the ferromagnetic susceptibility $\chi_F$. Anticipating
two-dimensional Ising critical behavior for which the susceptibility has a scale dimension of 7/4,  
we analyze the scaling collapse of $L^{-7/4} \chi_F$
\cite{S_note3}.
Figures\ \ref{scalferrornd}(c) and (d) show the scaling plots of the susceptibility data for
uncorrelated impurities at concentration $p=1/8$ and frustration parameters $g=0$ and 0.3, respectively.
As in the case of the Binder cumulants, the data collapse is of good quality. Values for $T_c$ and $\nu$
can be found by fitting the susceptibility to the expansion
\begin{equation}
\label{eq:suscfit}
L^{-7/4}\chi_{\rm F,S}(T,L) = f(x) = a_{0} + a_{1} x + a_{2} x^2 + \ldots.
\end{equation}
The resulting values are summarized in Table \ref{finaltable2}.
\begin{table}
\renewcommand*{\arraystretch}{1.2}
\begin{center}
\begin{tabular*}{13cm}{c |@{\extracolsep{\fill}} ccc|ccc}
\hline\hline
  ~  &\multicolumn{3}{c}{anticorrelated} & \multicolumn{3}{c}{uncorrelated} \\
$g$  & $T_{c}$&$\nu$ & $\bar{\chi}^{2}$ &$T_{c}$&$\nu$ &  $\bar{\chi}^{2}$ \\
\hline
$0$ &$1.7573(2)$ &$1.14(3)$& $0.64$ &$1.8031(2)$ &$1.10(2)$& $0.80$\\
$0.1$ &$1.4719(2)$ &$1.10(2)$ & $1.05$  &$1.5234(3)$&$1.13(4)$& $0.97$\\
$0.2$&$1.1720(2)$ &$1.12(3)$  & $0.96$ &$1.2287(2)$&$1.22(4)$& $0.72$ \\
$0.3$ &$0.8440(4)$ &$1.04(6)$ & $0.77$ &$0.9102(3)$&$1.18(4)$& $1.21$ \\
\hline\hline
\end{tabular*}
\caption{Critical temperatures $T_c$, effective correlation length exponents $\nu$, and reduced error sums $\bar\chi^2$
obtained from the scaling analysis of the ferromagnetic susceptibility $\chi_F$.
Results are shown for various values of the frustration parameter $g$ and dilution $p=1/8$
for both uncorrelated and anticorrelated impurities.}
\label{finaltable2}
\end{center}
\end{table}
They agree well with those from the analysis of the Binder cumulant. (For the effective exponent $\nu$,
the deviations are within one standard deviation; for $T_c$ they are within two standard deviations.)

\subsection*{S2.2 Stripe transition}

\begin{figure}
 \centerline{\includegraphics[width=0.75\textwidth,angle=0]{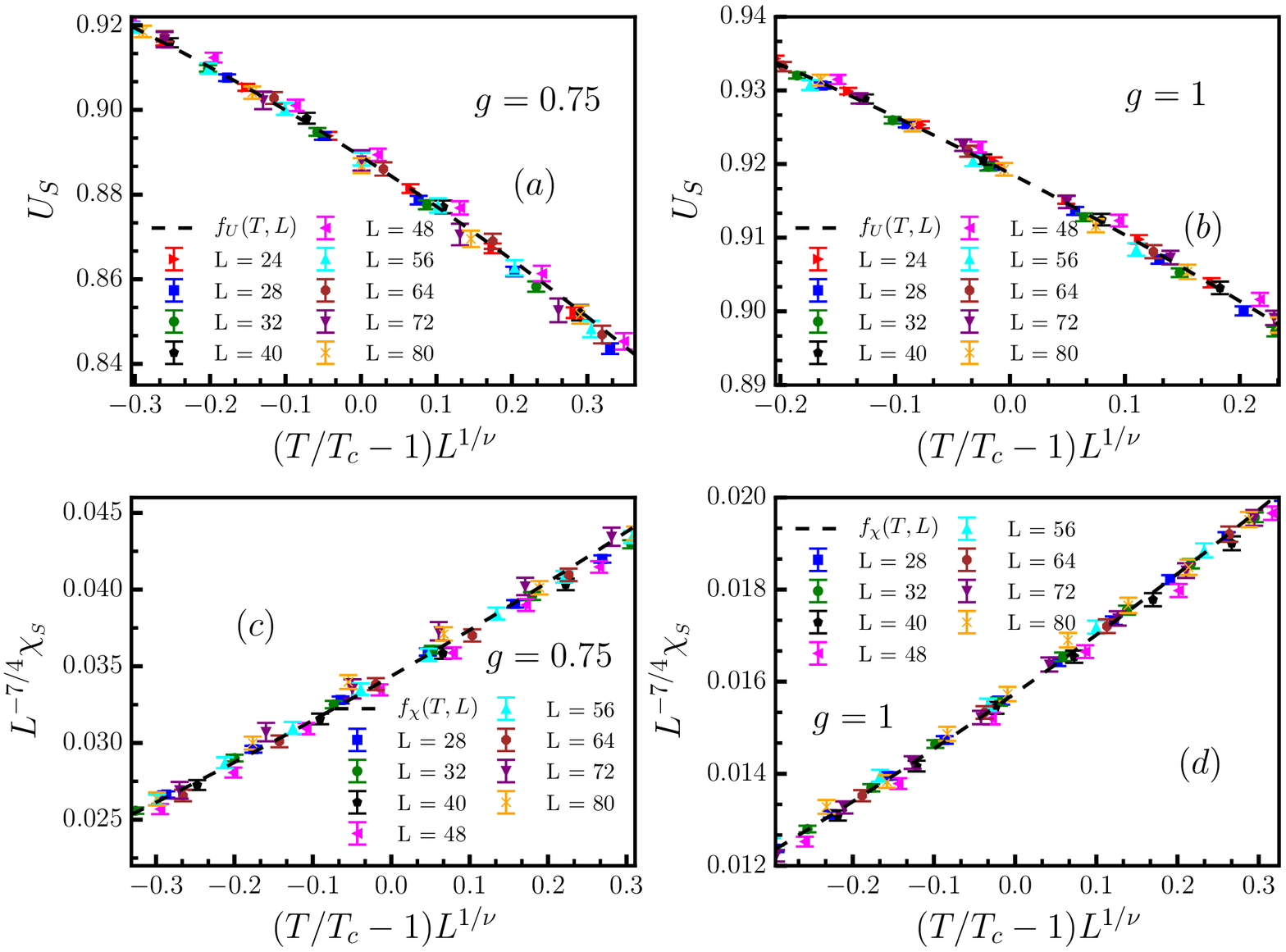}}
 \caption{Scaling plots of the stripe cumulant $U_{S}$ [panels (a) and (b)] and the stripe susceptibility $\chi_{S}$ 
 [panels (c) and (d)] for anticorrelated
  impurities of concentration of $p=1/8$ and frustration parameters $g=0.75$ and $g=1$. }
 \label{acstrsusscal}
 \end{figure}

The stripe-ordered to paramagnetic transition can be analyzed along the same lines as the ferromagnetic
transition above. Because uncorrelated impurities completely destroy the stripe phase, we only consider
perfectly anticorrelated impurities. Figure \ref{acstrsusscal} presents example scaling plots of the 
stripe Binder cumulant $U_S$ and the stripe susceptibility $\chi_S$ for impurity concentration $p=1/8$
and frustration parameters $g=0.75$ and $g=1$.
The values of $T_c$ and the correlation length exponent $\nu$ can again be determined from fits to
Eqs.\ (\ref{eq:binfit}) and (\ref{eq:suscfit}). The results are summarized in Table \ref{stripetable}.
\begin{table}
\renewcommand*{\arraystretch}{1.2}
\begin{center}
\begin{tabular*}{13cm}{c |@{\extracolsep{\fill}} ccc|ccc}
\hline\hline
  ~  &\multicolumn{3}{c}{Binder cumulant $U_S$} & \multicolumn{3}{c}{susceptibility $\chi_S$} \\
$g$ &  $T_{c}$ &$\nu$  & $\bar{\chi}^{2}$ &  $T_{c}$ &$\nu$  & $\bar{\chi}^{2}$\\
\hline
$0.60$   & $	0.70766(9)$ &$0.93(2)$&$0.92$ \\
$0.70$ & $0.9827(1)$ &$0.99(3)$ &$1.10$ & $0.9838(1)$ &$1.04(2)$  &$1.57$ \\
$0.75$ &$1.1020(1)$ &$1.00(2)$& $1.09$ &$1.1029(1)$ &$1.04(2)$ & $1.34$\\
$1$ &$1.6361(1)$ &$1.05(2)$& $1.09$ &$1.6362(1)$ &$1.07(1)$& $1.01$\\
\hline\hline
\end{tabular*}
\end{center}
\caption{Critical temperatures $T_c$, effective correlation length exponents $\nu$, and reduced error sums $\bar\chi^2$
obtained from the scaling analysis of the stripe Binder cumulant $U_S$ and the stripe susceptibility $\chi_S$.
Results are shown for various values of the frustration parameter $g$ and dilution $p=1/8$
for perfectly anticorrelated impurities.}
\label{stripetable}
\end{table}
In the undiluted, clean system, the stripe to paramagnetic transition is either of first-order (for $g<g^*\approx 0.67$)
or belongs to the Ashkin-Teller universality class (for $g>g^*$) \cite{kalz,kalzprb2012,arnab1,arnab2}). We have shown in the main
text that the first-order transition is rounded to a continuous one in the presence of anticorrelated impurities, 
as is expected from the Aizenman-Wehr theorem \cite{S_aizenmann}. Our results in Table \ref{stripetable} 
show that the critical exponent $\nu$ of the diluted system is close to the clean Ising value of unity for all
studied values of $g$. In particular, $\nu$ does not vary systematically with $g$ as would be
expected for the clean Ashkin-Teller universality class. The effects of disorder on the Ashkin-Teller
universality class were studied by Murthy \cite{S_Murthy} and Cardy \cite{S_Cardy} via a renormalization group analysis that 
predicted clean Ising critical behavior with universal logarithmic corrections just as in the
disordered Ising model. This was recently confirmed by large-scale simulations \cite{S_qiong}. 
As in the case of the ferromagnetic transition above, the system sizes in our present work are too small to 
extract logarithmic corrections. However, the effective $\nu$ values in
Table \ref{stripetable} are close to the clean two-dimensional Ising value of unity. We conclude that our results
are consistent with the critical behavior of the stripe transition belonging the disordered
Ising universality class.

%%%%%%%%%%%%%%%%%%%%%%%%%%%%%%%%%%%%%%%%%%%%%%%%%%%%%%%%%%%%%%%%%%%%%%%%%%%%%%%%%%%%%%%%%%%%%%%%%%%%%%%%%%%%%%%%%%%%%%%
\section*{S3. Domains}
%%%%%%%%%%%%%%%%%%%%%%%%%%%%%%%%%%%%%%%%%%%%%%%%%%%%%%%%%%%%%%%%%%%%%%%%%%%%%%%%%%%%%%%%%%%%%%%%%%%%%%%%%%%%%%%%%%%%%%%

As discussed in the main text, spinless impurities in the $J_1$-$J_2$ Hamiltonian create random fields for the nematic 
order parameter $\eta= \psi_x^2 -\psi_y^2$ which measures the local preference for vertical vs. horizontal stripes.
These random fields destroy the long-range stripe order via domain formation.
In order to image these domains, we define a local version of the nematic order parameter via 
$\eta_i = (\bar\psi_{i,x}^2-\bar\psi_{i,y}^2)$ where $\bar\psi_{i,x}$ and $\bar\psi_{i,y}$ are formed by averaging 
$\psi_{i,x} = S_i(-1)^{x_i}$, and $\psi_{i,y} = S_i(-1)^{y_i}$ over $2\times 2$ plaquette number $i$. 

\begin{figure}
\centering
\includegraphics[width=0.7\textwidth]{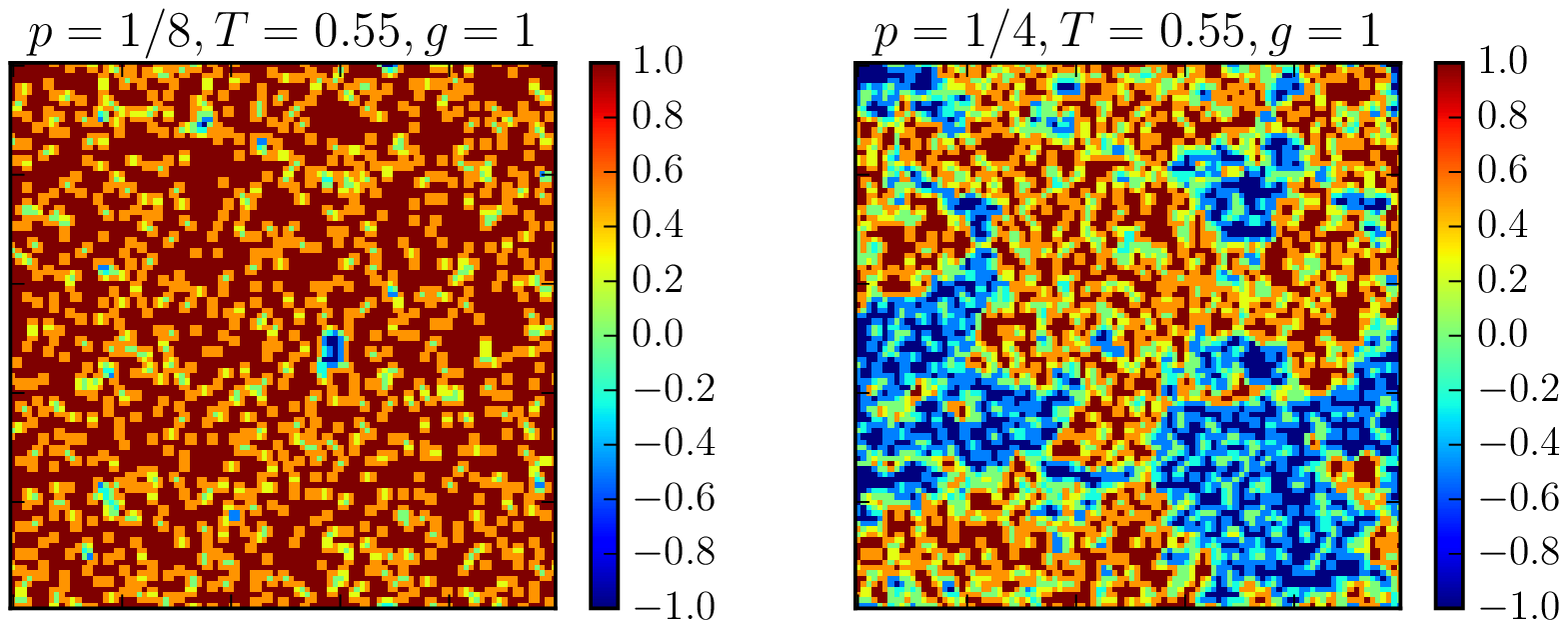}
\caption{Local nematic order parameter $\eta_i$ for each $2\times 2$ plaquette of a single system of 100x100 sites
for $T=0.55$, $g=1$, and uncorrelated impurities of concentration $p=1/8$ (left panel) and $p=1/4$ (right panel).}
\label{localopstr}
\end{figure}

Figure \ref{localopstr} illustrates the emergence of the domains in a system of linear size $L=100$ at $g=1$ and $T=0.55$ as we increase the concentration $p$ of impurities. For impurity concentration $p=1/8$, the local order parameter fluctuates only slightly,
i.e., the entire system belongs to a single domain. For the more disordered sample, $p=1/4$, the characteristic domain size has
fallen below the system size. The figure now shows random-field induced domain walls  percolating throughout the sample, 
thus leading to the destruction of long-range stripe order.


\begin{thebibliography}{11}
\bibitem{note1} We denote the reduced sum of squared errors of the fit (per degree
of freedom) by $\bar{\chi}^2$ to distinguish it from the susceptibility $\chi$.
\bibitem{note2} Strictly, this arguments holds at $T=0$ only.
At finite temperatures, random fields may be generated via entropic effects \cite{nattermann}.
However, at low temperatures they are expected to be extremely weak and likely unobservable
in experiments and simulations.
\end{thebibliography}
\begin{thebibliography}{11}
\bibitem{S_note3} As for the Binder cumulant, logarithmic corrections to the clean Ising behavior are expected. 
Our system sizes are too small, however, to reliably extract these logarithms.
\end{thebibliography}
\end{document}